\documentclass[12pt,pdftex]{article}  
\usepackage[top=3.5cm,bottom=3.5cm,left=2cm,right=2cm]{geometry}
\usepackage[dvipdfmx]{graphicx,color}
\usepackage{amssymb,amsmath}
\usepackage{slashed}
\usepackage{cite}
\usepackage{braket}
\usepackage{simplewick}
\usepackage[title]{appendix}
\usepackage{caption}

\usepackage{ascmac}
\usepackage{ulem}
\usepackage{latexsym}
\usepackage{pifont}          
\usepackage{geometry}
\usepackage{bm}

\newcommand{\n}{\nonumber}

\newcommand{\mr}[1]{\mathrm{#1}}

\newcommand{\h}[1]{\hspace{#1}}
\newcommand{\aln}[1]{\begin{align}#1\end{align}}

\newcommand{\s}[1]{\slashed{#1}}

\begin{document}

\title{
\vbox{
\baselineskip 14pt
\hfill \hbox{\normalsize KUNS-2728
}} \vskip 1cm
\bf \Large  A novel regularization of chiral gauge theory
\vskip 0.5cm
}
\author{
Yu~Hamada\thanks{E-mail: \tt yu.hamada@gauge.scphys.kyoto-u.ac.jp},~
Hikaru~Kawai\thanks{E-mail: \tt hkawai@gauge.scphys.kyoto-u.ac.jp}, and
Katsuta~Sakai\thanks{E-mail: \tt katsutas@gauge.scphys.kyoto-u.ac.jp} 
\bigskip\\
\it \normalsize
 Department of Physics, Kyoto University, Kyoto 606-8502, Japan\\
\smallskip
}
\date{}

\setcounter{page}{0}
\maketitle\thispagestyle{empty}
\abstract{\normalsize
We propose a novel gauge-invariant regularization for the perturbative chiral gauge theory. 
Our method consists of the two ingredients: use of the domain-wall fermion to describe a chiral fermion with Pauli-Villars regulators and application of the dimensional regularization only to the gauge field.
This regularization is implemented in the Lagrangian level, unlike other gauge-invariant regularizations (eg. the covariant regularizations).
We show that the Abelian (fermion number) anomaly is reproduced correctly in this formulation.
We also show that once we add the counter terms to the full theory, then the renormalization in the chiral gauge theory is automatically achieved.

}
\newpage
\tableofcontents
\newpage
\section{Introduction}
\label{sec: introduction}
\quad
Chiral gauge theory is one of the most essential theories to describe nature, since the Standard Model includes the chiral sector $SU(2)_L\times U(1)_Y$. 
Among various subjects around it, its (non-) perturbative regularization has been attracting much interest. 
It is mainly because the theory has an anomaly and no regulator exists that maintain the gauge invariance. 
If one na\"ively regularizes chiral gauge theory by applying the ordinary scheme such as the dimensional regularization or Pauli-Villars (PV) regularization, one obtains the anomaly from one-loop Feynman diagrams. 
In that case, however, one has to add local finite counter terms to the action in order to recover the gauge symmetry, even when the theory is free from anomaly. 
We call the counter terms the fake anomaly. 
The computation of the necessary counter terms is quite complicated in general, and it is natural to ask whether there is some regularization scheme that yields no fake anomaly. 
At the level of the one-loop Feynman diagram, it has been known that the covariant regularizations serve as such schemes. 
They are equivalent to regularizations of the change of the fermion measure covariantly \cite{Fujikawa:1979ay}, 
and the diagram gives only the covariant anomaly.\footnote{Although the consistent and covariant anomalies differ by the Bardeen-Zumino current \cite{Bardeen:1984}, 
it vanishes when the theory is free from any gauge anomaly. Therefore one needs no counter term in anomaly-free cases.} 
On the other hand, there is no such formulation known at the Lagrangian level; the regularized Lagrangian to realize some covariant regularization remains to be constructed. 
So far it has been partially achieved in various works. 
For example, the generalized PV regularization \cite{Frolov:1992ck} is one of the sophisticated formulation, 
which regularizes the chiral gauge theory covariantly as long as the gauge anomaly is absent. It also reproduces the Abelian anomaly with the correct coefficient \cite{Aoki:1993fz}.  
It has been shown that when one applies the generalized PV regularization, 
the regularized contribution can be regarded as a sort of the covariant regularization \cite{Fujikawa:1994np}. 
However, some improvement is still required since this scheme fails to regularize the parity-odd 
contribution. 

On the other hand, there is a theory called the domain-wall (DW) fermion \cite{Kaplan:1992bt}, 
where a chiral fermion is induced from a higher dimensional Dirac fermion with a topological defect in its mass. 
Its relationship to the generalized PV regularization was discussed in \cite{Narayanan:1992wx}. 
The theories have been thoroughly investigated in the context of the lattice gauge theory \cite{Narayanan:1993sk,Narayanan:1993ss,Slavnov:1993pg,Aoki:1993rg,Furman:1994ky,Shamir:1993zy,Golterman:1993th,Kikukawa:2001mw,Fukaya:2016ofi,Grabowska:2015qpk,Grabowska:2016bis,Makino:2016auf,Makino:2017pbq,Okumura:2016dsr,Hamada:2017tny}. 
In particular, it has been proposed in \cite{Grabowska:2015qpk, Grabowska:2016bis} that the lattice regularization of chiral gauge theory can be realized by combining the DW fermion and the gradient flow of the gauge field \cite{Luscher:2010iy, Luscher:2011bx}. 

Although the DW fermion was originally discussed in the context of the lattice gauge theory, it should be useful to pursue a perturbative formulation as well. 
In this paper, we propose a regularization with which no fake anomaly emerges by defining chiral gauge theory through the DW fermion. 
The significant point is that the theory is vector-like and its regularized version is expected to induce chiral gauge theory without fake anomaly. 
One might guess that the problem is readily solved by combining the DW fermion with the dimensional regularization. 
However, it is not the case for the following reason. For example, consider a $(4+\varepsilon+1)$-dimensional DW fermion. 
Since its kinetic term respects the $(4+\epsilon+1)$-dimensional Lorentz invariance, the $5$-th and $\varepsilon$-th components of loop momenta $p$ and 
the corresponding gamma matrices enter the loop integrand as a combination $\gamma_{5}p_{5}+\slashed{p}_\varepsilon$. It can be rewritten as $\gamma_{5}(p_{5}+\gamma_5\slashed{p}_\varepsilon)$. 
Here, $\gamma_5$ is the chiral operator with respect to the 4-dimensional induced chiral gauge theory. 
Because $\gamma_5\slashed{p}_\varepsilon$ commutes with all the 4-dimensional gamma matrices, we can replace it with its eigenvalues $\pm(p_\varepsilon^2)^{1/2}$ in each eigenspace to find that the integrand depends on $p_5\pm(p_\varepsilon^2)^{1/2}$. 
As a result, $p_\varepsilon$, the integration over that would lead to the finiteness of the loop integration in the usual dimensional regularization, is absorbed in the integration over $p_5$. 
It leaves the overall loop integration unregularized. In order to circumvent this difficulty around the dimensional regularization and the DW fermion, 
we propose the partially dimensional regularization (PDR), where we apply the dimensional regularization to not all of field contents. 
In the case of the DW fermion, the dimensional regularization is applied only to the gauge field, while fermions are regularized with the PV regularization. 
As will be demonstrated in this paper, this PDR might be useful to calculate the effective action.

This paper is organized as follows. 
In Sec.\ref{sec: DW}, we briefly review the DW fermion and the induced chiral fermion loops on the domain-wall. 
There the vacuum polarization was discussed as an explicit example to see how we regularized fermion loops with the PV regularization. 
Then in Sec.\ref{sec: chiral_anomaly}, we show that this scheme gives the consistent anomaly in the 2-dimensional case. 
In Sec.\ref{004700_18May18}, we present how the PDR works as a regularization with an example of the 4-dimensional Yukawa theory. 
It is shown that the non-conservation of $\varepsilon$-th components of momenta results in the regularization of loop integrations. 
This mechanism slightly differs from the usual dimensional regularization. 
In Sec.\ref{143839_21May18}, we analyze the case where the PDR is applied to the 5-dimensional DW fermion. 
In the theory the induced chiral gauge theory is 4-dimensional. We investigate the self-energy of the fermion and discuss the renormalization. 
Finally in Sec.\ref{sec: summary} we give the summary and several open questions.

\section{A Brief Review of the Domain-Wall Fermions}
\label{sec: DW}
\quad
Let us have a brief review of some results in our previous work \cite{Hamada:2017tny}, where we have stated that one can regularize a chiral fermion loop with a sort of PV fields. 
Although the main topic of the work was related to the gradient flow, the discussion around the regularization also holds in the case where the flow is switched off.

\begin{figure}
\begin{center}
\includegraphics[width=8cm,clip]{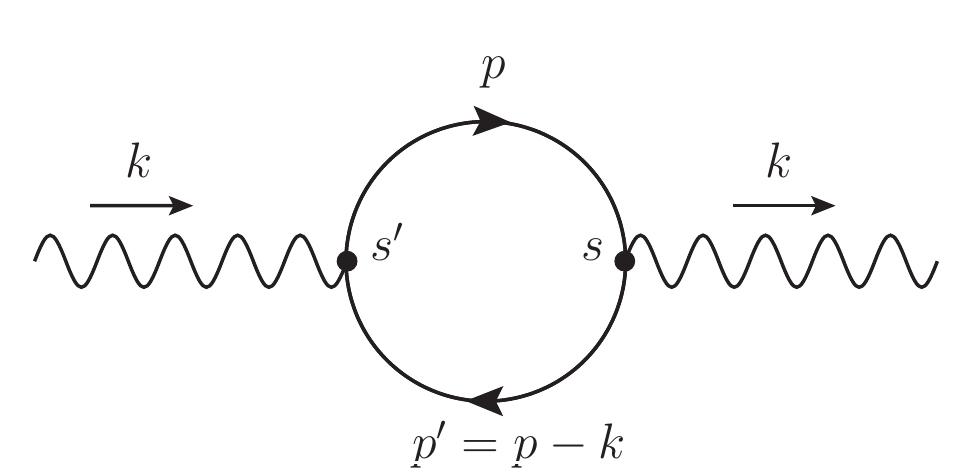}
\caption{The diagram of the vacuum polarization involving a fermion loop. Although we compute it as a $(2d+1)$-dimensional diagram, 
one can integrate out the $s$ and $s'$-dependence of the localized mode to obtain an effectively $2d$-dimensional quantity. 
}
\label{fig: rev_vac_pol}
\end{center}
\end{figure}

Consider the $(2d+1)$-dimensional action for a DW fermion: 
\aln{
S=&\int \frac{d^{2d}p}{(2\pi)^{2d}}ds\Bigl[\bar{\psi}(-p,s)\left(i\slashed{p}+\slashed{A}+\gamma_5\partial_s-\epsilon(s)M\right)\psi(p,s)\nonumber\\
&~~~~~~~~~~~~~~~~~+\bar{\phi}(-p,s)\left(i\slashed{p}+\slashed{A}+\gamma_5\partial_s+M\right)\phi(p,s)\Bigr]\nonumber\\
&+\int \frac{d^{2d}p}{(2\pi)^{2d}}\frac{1}{4}F_{\mu\nu}(-p)F_{\mu\nu}(p), 
\label{eq: rev_DW_action}
}
where we have represented the fields in the momentum space with respect to $2d$ directions, and in the coordinate space with respect to the $(2d+1)$th direction. 
We have denoted the coordinate for this direction as $s$.  
$\psi$ is the ordinary DW fermion, while $\phi$ is an auxiliary bosonic spinor field which is necessary for subtracting undesirable bulk contributions. 
$\gamma_5$ is the gamma matrix corresponding to the $s$-direction, or the chiral operator from the viewpoint of the $2d$-dimensional domain-wall.\footnote{Although we consider a $(2d+1)$ -dimensional theory we have assigned the subscript ``five'' to the matrix, since we would like to eventually describe a four-dimensional chiral gauge theory and the matrix will correspond to the usual chiral operator.} 
$F_{\mu\nu}$ is a $2d$-dimensional gauge field strength, and we define its couplings to $\psi$ and $\phi$ through copying it along the $s$-direction. 
It implies $A_{2d+1}=0$. 
The propagators for $\psi$ and $\phi$ are obtained from the action (\ref{eq: rev_DW_action}), and are given by 
\aln{
G_{\psi}(p,s,s')=\left\{
\begin{array}{ll}
S^{(+)}+D^{(+)}&(0<s,s')\\
D^{(-+)}&(s<0<s')\\
D^{(+-)}&(s'<0<s)\\
S^{(-)}+D^{(-)}&(s,s'<0)
\end{array}\right.,~~~~G_{\phi}(p,s,s')=S^{(-)}, \label{154523_24May18}
}
where
\aln{
S^{\pm}(p,s,s')=&-\left[\theta(s-s')\frac{1}{2}\left(\frac{i\slashed{p}\pm M}{\sqrt{p^2+M^2}}-\gamma_5\right)e^{-(s-s')\sqrt{p^2+M^2}}\right.\nonumber\\
&~~~~~~\left.+\theta(s'-s)\frac{1}{2}\left(\frac{i\slashed{p}\pm M}{\sqrt{p^2+M^2}}+\gamma_5\right)e^{-(s'-s)\sqrt{p^2+M^2}}\right],\label{154558_24May18}\\
D^{\pm}(p,s,s')=&\mp M\frac{i\slashed{p}}{p^2}\frac{1}{2}\left(\frac{i\slashed{p}\pm M}{\sqrt{p^2+M^2}}\pm\gamma_5\right)e^{\mp(s+s')\sqrt{p^2+M^2}},\label{154624_24May18}\\
D^{\mp\pm}(p,s,s')=&\mp\sqrt{p^2+M^2}\frac{i\slashed{p}}{p^2}\frac{\gamma_5}{2}\left(\frac{i\slashed{p}\pm M}{\sqrt{p^2+M^2}}\pm\gamma_5\right)e^{\pm(s-s')\sqrt{p^2+M^2}}.\label{154629_24May18} 
}
Here $S^{(\pm)}$ are the usual propagators for $(2d+1)$-dimensional fermions with their masses $\pm M$. 
On the other hand, $D^{(\pm)}$ and $D^{(\mp\pm)}$ are ones for the massless modes localized on the domain-wall, as is seen from the factors of exponential. 
They correspond to the $2d$-dimensional chiral fermion. 
Using the propagators, the vacuum polarization with an external momentum $k_\mu$ (Fig.(\ref{fig: rev_vac_pol})) is calculated as
\aln{
\int \frac{d^{2d}p}{(2\pi)^{2d}}\int_{-\infty}^{+\infty}\!ds\int_{-\infty}^{+\infty}\!ds'\,\Pi_{\mu\nu}
=&\int \frac{d^{2d}p}{(2\pi)^{2d}}\int_{0}^{+\infty}\!ds\int_{0}^{+\infty}\!ds'\,\Pi_{\mu\nu}^{blk}+\int \frac{d^{2d}p}{(2\pi)^{2d}}\,\Pi_{\mu\nu}^{DW},
\label{eq: rev_Pi}
}
where
\aln{
\Pi_{\mu\nu}^{blk}=&\mathrm{Tr}[S^{(+)}\gamma_\mu S^{(+)\prime}\gamma_\nu]-\mathrm{Tr}[S^{(-)}\gamma_\mu S^{(-)\prime}\gamma_\nu],\\
\Pi_{\mu\nu}^{DW}=&\int_{-\infty}^{0}\! ds\int_{-\infty}^{0}\! ds'\Bigl\{\mathrm{Tr}[D^{(-)}\gamma_\mu D^{(-)\prime}\gamma_\nu]
+\mathrm{Tr}[D^{(-)}\gamma_\mu S^{(-)\prime}\gamma_\nu]+\mathrm{Tr}[S^{(-)}\gamma_\mu D^{(-)\prime}\gamma_\nu]\Bigr\}\nonumber\\
&+\int_{-\infty}^0\! ds\int_0^{+\infty}\! ds'\Bigl\{\mathrm{Tr}[D^{(-+)}\gamma_\mu D^{(+-)\prime}\gamma_\nu]-\mathrm{Tr}[S^{(-)}\gamma_\mu S^{(-)\prime}\gamma_\nu]\Bigr\}\nonumber\\
&+\int_0^\infty \! ds\int_{-\infty}^0\! ds'\Bigl\{\mathrm{Tr}[D^{(+-)}\gamma_\mu D^{(-+)\prime}\gamma_\nu]-\mathrm{Tr}[S^{(-)}\gamma_\mu S^{(-)\prime}\gamma_\nu]\Bigr\}\nonumber\\
&+\int^{+\infty}_{0}\! ds\int^{+\infty}_{0}\! ds'\Bigl\{\mathrm{Tr}[D^{(+)}\gamma_\mu D^{(+)\prime}\gamma_\nu]
+\mathrm{Tr}[D^{(+)}\gamma_\mu S^{(+)\prime}\gamma_\nu]+\mathrm{Tr}[S^{(+)}\gamma_\mu D^{(+)\prime}\gamma_\nu]\Bigr\}. 
\label{eq: rev_vac_pol}
}
Here the arguments of the factors with primes are $(p',s',s)$ with $p'=p-k$, while those without primes are $(p,s,s')$. 
$\Pi_{\mu\nu}^{DW}$ and $\Pi_{\mu\nu}^{blk}$ represent the contributions which is and is not localized on the domain-wall, respectively. 
$\Pi_{\mu\nu}^{blk}$ vanishes in the region $s<0$, where the massive modes of $\psi$ and $\phi$ cancel each other. 
Furthermore, one can easily find out that $\Pi_{\mu\nu}^{blk}$ actually vanishes even in the region $s>0$ for the following observation. 
The parity-even contributions of $\psi$ and $\phi$ to $\Pi_{\mu\nu}^{blk}$ are canceled, since they depend on even powers of $M$ and take the same forms except for the overall sign. 
On the other hands, the parity-odd contributions would yield the Charn-Simons term, which vanishes in our setup where $A_\mu$ is independent of $s$ and $A_{2d+1}=0$. 
Therefore, one concludes that 
\aln{
\Pi_{\mu\nu}^{blk}\equiv 0. 
}
At the same time, however, $\Pi_{\mu\nu}^{DW}$ has a contribution of the massive modes $S^{\pm}$ as appears in Eq.(\ref{eq: rev_vac_pol}). 
It is the result of the interactions to the modes localized on the domain-wall as in Eq.(\ref{eq: rev_vac_pol}). 

After some calculation we obtain the following result: 
\aln{
\Pi_{\mu\nu}^{DW}=&\frac{1}{2}\biggl\{\mathrm{Tr}\left[\frac{i\slashed{p}}{p^2}\gamma_\mu\frac{i\slashed{p}'}{p^{\prime\,2}}\gamma_\nu\right]
-\mathrm{Tr}\left[\frac{i\slashed{p}+M}{p^2+M^2}\gamma_\mu\frac{i\slashed{p}'+M}{p^{\prime\,2}+M^2}\gamma_\nu\right]\biggr\}\nonumber\\
&+\frac{1}{4}\mathrm{Tr}\left[\frac{i\slashed{p}}{p^2}\hat{\gamma}_5\,\gamma_\mu\frac{i\slashed{p}'}{p^{\prime\,2}}\gamma_\nu\right]
+\frac{1}{4}\mathrm{Tr}\left[\frac{i\slashed{p}}{p^2}\gamma_\mu\frac{i\slashed{p}'}{p^{\prime\,2}}\hat{\gamma}_5'\,\gamma_\nu\right]
\label{eq: rev_vac_pol2}
}
with
\aln{
\hat{\gamma}_5=2\left(1-\frac{\sqrt{p^2+M^2}\left(\sqrt{p^2+M^2}\sqrt{p^{\prime\,2}+M^2}+M^2\right)}{\sqrt{p^{\prime\,2}+M^2}\left(\sqrt{p^2+M^2}+\sqrt{p^{\prime\,2}+M^2}\right)^2}\right)
\frac{M}{\sqrt{p^2+M^2}}\gamma_5,
\label{eq: rev_deformed_chiral}
}
and $\hat{\gamma}_5'$ being the same expression with $p$ and $p'$ exchanged. 
We emphasize that Eq.(\ref{eq: rev_vac_pol2}) is a nontrivial result of the combination and cancellation between terms from each lines in Eq.(\ref{eq: rev_vac_pol}).
In particular, the first line of Eq.(\ref{eq: rev_vac_pol2}) contains no root factor such as $\sqrt{p^2+M^2}$, and is a local expression. 
It is regarded as the parity-even contribution with a single usual PV field. 
The leading divergent part is thus eliminated. 
On the contrary, the second line is highly non-local expression, 
and it can be interpreted as fermion loops with a deformed chiral operator Eq.(\ref{eq: rev_deformed_chiral}) inserted either of the two vertices. 
When the loop momentum is sufficiently large, {\it i.e.} $p,p'\gg M$, the deformed operator is roughly evaluated as
\aln{
\hat{\gamma}_5,\,\hat{\gamma}_5'\sim\frac{M}{\sqrt{p^2}}\,\gamma_5.
}
This means that the second line of Eq.(\ref{eq: rev_vac_pol2}) contains a suppression factor of $p^{-1}$ on the loop divergence. 
Moreover, the factor makes it possible to expand the contribution with respect to $M^2/p^2$ and $M^2/p^{\prime\,2}$, to obtain a series of $M$ with the odd-order. 
By expanding the entire $\Pi_{\mu\nu}^{DW}$, we obtain the following form:
\aln{
\Pi_{\mu\nu}^{DW}=\sum_{n\geq1}a_n\,M^{2n}\,\mathrm{Tr}[A_n]+\sum_{n\geq1}b_n\,M^{2n-1}\,\mathrm{Tr}[B_n]. 
\label{eq: rev_vac_pol_expanded}
}
Here $a_n$ and $b_n$ are numerical coefficients, $A_n$ and $B_n$ are matrices of $O(p^{-2n-2})$ and $O(p^{-2n-1})$, respectively. 
Each of $B_n$ includes one $\gamma$, and they represent the parity-odd part. 
In particular, in the case of $d=1$ and the resulting gauge theory is anomalous, they represent the anomaly.

In order to regularize $\Pi_{\mu\nu}^{DW}$, it is sufficient to introduce ``PV pairs'' and cancel the first few terms in Eq.(\ref{eq: rev_vac_pol_expanded}). 
Therefore, the action to consider is 
\aln{
S\rightarrow\int \frac{d^{2d}p}{(2\pi)^{2d}}ds\Bigl[&\bar{\psi}(-p,s)\left(i\slashed{p}+\slashed{A}+\gamma_5\partial_s-\epsilon(s)M\right)\psi(p,s)\nonumber\\
&+\bar{\phi}(-p,s)\left(i\slashed{p}+\slashed{A}+\gamma_5\partial_s+M\right)\phi(p,s)\nonumber\\
&+\sum_{i=1}^m\sum_{r=1}^{|C_i|}\bar{\psi}_{i,r}(-p,s)\left(i\slashed{p}+\slashed{A}+\gamma_5\partial_s-\epsilon(s)M_i\right)\psi_{i,r}(p,s)\nonumber\\
&+\sum_{i=1}^m\sum_{r=1}^{|C_i|}\bar{\phi}_{i,r}(-p,s)\left(i\slashed{p}+\slashed{A}+\gamma_5\partial_s+M_i\right)\phi_{i,r}(p,s)\Bigr]\nonumber\\
+\int \frac{d^{2d}p}{(2\pi)^{2d}}\frac{1}{4}&F_{\mu\nu}(-p)F_{\mu\nu}(p), 
\label{eq: rev_action_reg}
} 
where the masses of the additional pairs, $\psi_{i,r}$ and $\phi_{i,r}$, are denoted as $M_i$. 
For each $i$, all of $\psi_{i,r}$'s are simultaneously either fermionic or bosonic, and the corresponding $\phi_{i,r}$'s follow the opposite statistics. 
We have introduced $\sum_{i=1}^m|C_i|$ PV pairs in total. 
We take the sign of an integer $C_i$ positive when $\psi_{i,r}$'s are fermionic.   
The PV pairs yield the same form of contributions to the vacuum polarization as the original pair, changing $\Pi_{\mu\nu}^{(DW)}$ as
\aln{
\Pi_{\mu\nu}^{DW}=\sum_{i=0}^mC_i\left(\sum_{n\geq1}a_n\,M_i^{2n}\,\mathrm{Tr}[A_n]+\sum_{n\geq1}b_n\,M_i^{2n-1}\,\mathrm{Tr}[B_n]\right) 
}
with $M_0=M$ and $C_0=1$. Therefore, the conditions to regularize $\Pi_{\mu\nu}^{DW}$ are summarized as follows:
\aln{
\sum_{i=0}^mC_iM_i&=M+\sum_{i=1}^mC_iM_i=0,\nonumber\\
\sum_{i=0}^mC_iM_i^2&=M^2+\sum_{i=1}^mC_iM_i^2=0,\nonumber\\
\sum_{i=0}^mC_iM_i^3&=M^3+\sum_{i=1}^mC_iM_i^3=0,\nonumber\\
&~~\vdots
\label{eq: rev_cond}
}
Note that the zeroth-order condition $\sum_{i=0}^mC_i=0$ is not necessary because it corresponds to elimination of the leading divergence, 
which have been removed automatically in Eq.(\ref{eq: rev_vac_pol2}). 
Rather, in order to prevent the PV pairs from generating extra massless modes, we require
\aln{
\sum_{i=1}^mC_i=0.
\label{eq: rev_cond2}
}
Eqs.(\ref{eq: rev_cond}) and Eq.(\ref{eq: rev_cond2}) can be solved with respect to $C_i$ and $M_i$ when we introduce a sufficiently number of the PV pairs. 
The most important point of the above analysis is that the parity-odd part has a non-locally deformed chiral operator, 
and that it enables us to regularize that part in the same manner as the ordinary PV regularization. 
Therefore, we have obtained the way to regularize the vacuum polarization even in the two-dimensional theory is anomalous. 
Note that it works even for the two-dimensional theory with anomaly. 
The general loop diagrams are also regularized by Eq.(\ref{eq: rev_action_reg}).

Finally, we briefly comment on why the present regularization yields the gauge-invariant result in the anomaly-free case, as mentioned in \cite{Hamada:2017tny}. 
If we take the limit $M\to\infty$, Eq.(\ref{eq: rev_vac_pol2}) takes the form of 
\aln{
\Pi_{\mu\nu}^{DW}\to&~\frac{1}{2}\mathrm{Tr}\left[\frac{i\slashed{p}}{p^2}\gamma_\mu\frac{i\slashed{p}'}{p^{\prime\,2}}\gamma_\nu\right]
+\frac{1}{4}\mathrm{Tr}\left[\frac{i\slashed{p}}{p^2}\gamma_5\,\gamma_\mu\frac{i\slashed{p}'}{p^{\prime\,2}}\gamma_\nu\right]
+\frac{1}{4}\mathrm{Tr}\left[\frac{i\slashed{p}}{p^2}\gamma_\mu\frac{i\slashed{p}'}{p^{\prime\,2}}\gamma_5\,\gamma_\nu\right]\nonumber\\
=&~\frac{1}{2}\biggl(\mathrm{Tr}\left[\frac{i\slashed{p}}{p^2}\frac{1+\gamma_5}{2}\,\gamma_\mu\frac{i\slashed{p}'}{p^{\prime\,2}}\gamma_\nu\right]
+\mathrm{Tr}\left[\frac{i\slashed{p}}{p^2}\gamma_\mu\frac{i\slashed{p}'}{p^{\prime\,2}}\frac{1+\gamma_5}{2}\,\gamma_\nu\right]\biggr). 
\label{eq: rev_cov}
}
Note that if one regularizes Eq.(\ref{eq: rev_cov}), it is regarded as a covariant regularization, 
a prescription where one regularizes one-loop diagrams with only one projection operator inserted at some vertex. 
Then it is natural that the regularized theory Eq.(\ref{eq: rev_action_reg}) gives the gauge-invariant parity-even part. 
Note that Eq.(\ref{eq: rev_cov}) includes the average over the points where the projection operator is inserted. The bose symmetry is thus automatically maintained. 
These facts imply that the theory has no fake anomaly.\footnote{The present regularization is specifically interesting because it regularizes the parity-even part by PV regularization, 
while it deforms non-locally the parity-odd part with the same PV mass.} 
However, as is mentioned above, the form of Eq.(\ref{eq: rev_vac_pol2}) has been obtained as the result of all the relevant calculation 
(integration over $s$-direction, combination and cancellation of the terms).
Although the DW fermion indeed realizes a good regularization, the relationship to the covariant regularizations still remains to be figured out.

\section{Chiral Anomaly}
\label{sec: chiral_anomaly}
As is well known, chiral gauge theory has the abelian anomaly (fermion number anomaly).
In this section, we derive the anomaly in our regularization.
We find that the gauge invariant fermion number current is defined as
\begin{align}
 J_\mu(x)&\equiv \int ds \Big[ \bar{\psi}(x,s)\gamma_\mu\psi(x,s) +  \bar{\phi} (x,s)\gamma_\mu \phi(x,s) \n \\
 &+ \sum_{i=1}^m\sum_{r=1}^{|C_i|} \left[ \bar{\psi}_{i,r}(x,s)\gamma_\mu \psi_{i,r}(x,s) + \bar{\phi}_{i,r}(x,s)\gamma_\mu \phi_{i,r}(x,s)\right] \Big] \label{014202_1May18},
\end{align}
which is the Noether current for the $U(1)$ rotation of the DW fermion, subtracting field, and all the regulator fields.
We will show that this current reproduces the correct anomaly.

\subsection{Evaluation of anomaly}
In the following, we evaluate the expectation value of the current Eq.\eqref{014202_1May18} in the presence of the background gauge field $A_\mu$.
We focus on the case of the two-dimensional Abelian chiral gauge theory.
In this case, we need not introduce the PV pairs because the only divergent part in the fermion loop is already subtracted with $\phi$. 

\begin{figure}[tbp]
  \begin{center}
  \includegraphics[width=0.35\textwidth]{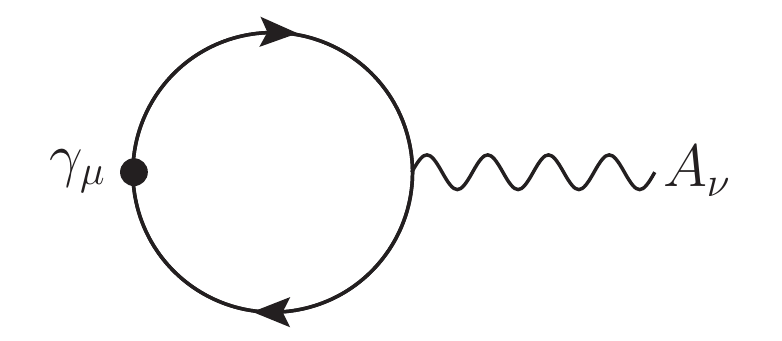}
  \caption{A diagram representing the anomalous current}
\label{025514_21May18}
  \end{center}
 \end{figure}

The Feynman diagram corresponding to the current is shown in Fig.\ref{025514_21May18}.
This expectation value can be expressed in terms of the vacuum polarization Eq.\eqref{eq: rev_Pi} as follows:
\begin{align}
 \braket{J_\mu(k)}_A &= \int \frac{d^2p}{(2\pi)^2} \int ds \int ds' ~\Pi_{\mu\nu} A_\nu(k) \\
 &=  \int \frac{d^2p}{(2\pi)^2} ~\Pi_{\mu\nu}^{DW} A_\nu(k) .\label{181742_8May18}
\end{align}
Recall that $\Pi_{\mu\nu}^{blk}$ vanishes as stated in the previous section. 
Using the fact that the parity-even part of $\Pi_{\mu\nu}^{DW}$ is transverse, the divergence of the current is expressed as 
\begin{align}
 k_\mu\braket{J_\mu}_A &= \int \frac{d^2 p}{(2\pi)^2}~k_\mu  \Pi_{\mu\nu}^{DW} A_\nu\\
 & =\frac{1}{4} A_\nu \int \frac{d^2 p}{(2\pi)^2}~\mr{Tr}\left[\frac{1}{i\s{p}}\hat{\gamma_5}\s{k}\frac{1}{i\s{p}'}\gamma_\nu + \frac{1}{i\s{p}}\s{k}\frac{1}{i\s{p}'}\hat{\gamma_5}'\gamma_\nu \right]  \\
 &=\frac{-1}{4}A_\nu \int \frac{d^2 p}{(2\pi)^2}~(f(p,p')+f(p',p))\mr{Tr}\left[\left(\frac{1}{\s{p}}-\frac{1}{\s{p}'}\right)\gamma_5 \gamma_\nu \right] ,
\end{align}
where
\begin{equation}
 f(p_1,p_2)\equiv 2\left(1-\frac{\sqrt{(p_1)^2+M^2}\left(\sqrt{(p_1)^2+M^2}\sqrt{(p_2)^2+M^2}+M^2\right)}{\sqrt{(p_2)^2+M^2}\left(\sqrt{(p_1)^2+M^2}+\sqrt{(p_2)^2+M^2}\right)^2}\right) \frac{M}{\sqrt{(p_1)^2+M^2}}.
\end{equation}
%
%

By expanding $f$ as 
\begin{equation}
 f(p,p') \sim f(p',p) \sim 2\left( \frac{3}{4}- \frac{M^2}{4(p^2+M^2)}\right)\frac{M}{\sqrt{p^2+M^2}} + \mathcal{O}(k^2/M^2)\label{133751_8May18} 
\end{equation}
and taking the limit $k^2/M^2 \to 0$, we obtain
\begin{align}
 \int \frac{d^2 p}{(2\pi)^2}~ k_\mu \Pi_{\mu\nu}^{DW} & \to  -  \int \frac{d^2 p}{(2\pi)^2}  \left( \frac{3}{4} - \frac{M^2}{4(p^2+M^2)}\right)\frac{M}{\sqrt{p^2+M^2}} \mr{Tr}\left[\left(\frac{1}{\s{p}}-\frac{1}{\s{p}'}\right) \gamma_5 \gamma_\nu\right] \\
 &=   \int \frac{d^2 p}{(2\pi)^2} \left( \frac{3}{4} - \frac{M^2}{4(p^2+M^2)}\right)\frac{M}{\sqrt{p^2+M^2}} \mr{Tr}\left[\frac{\s{p}'}{p'^2} \gamma_5 \gamma_\nu\right] .\label{141530_8May18}
\end{align}
Here the term propotional to $1/\s{p}$ has vanished in the last line because it is odd in $p$.

By using the Feynman parametrization
\begin{equation}
 \frac{1}{\left[p^2+M^2\right]^n } \frac{1}{p'^2} = n \int_0^1 dx ~\frac{(1-x)^{n -1}}{\left[ (1-x)(p^2+M^2)+x(p - k)^2 \right]^{1+n}}
\end{equation}
for $n=\frac{1}{2}$ and $\frac{3}{2}$, Eq.\eqref{141530_8May18} can be rewritten as the following expression:
\begin{align}
 \frac{3M}{8}  \int \frac{d^2 q}{(2\pi)^2} \int_0^1 dx ~\left(\frac{(1-x)^{-\frac{1}{2}}}{\left[q^2 + \Delta\right]^{3/2}}-\frac{M^2(1-x)^{\frac{1}{2}}}{\left[q^2 + \Delta\right]^{5/2}} \right)~\mr{Tr}\left[\s{p}'\gamma_5\gamma_\nu\right]
\end{align}
with
\begin{equation}
 q \equiv p-xk, \h{3em}\Delta\equiv x(1-x)k^2 + (1-x)M^2.
\end{equation}
Its integration over $q$ leads to the following form:
\begin{align}
 & \frac{3M}{16\pi}~ \mr{Tr}\left[\s{k}\gamma_5\gamma_\nu\right]\int_0^1 dx ~\left[\left(\frac{1-x}{\Delta} \right)^{\frac{1}{2}}-\frac{M^2}{3}\left(\frac{1-x}{\Delta}\right)^{\frac{3}{2}}\right]\\
 &\to -\frac{1}{4\pi } \frac{M}{|M|}k_\mu \epsilon_{\mu\nu} \h{2em} (\mr{as}~~ k/M \to 0).
\end{align}

The resulting expression for the divergence of the current is  
\begin{equation}
 \braket{k_\mu J_\mu}_A = \frac{-1}{4\pi}\frac{M}{|M|}\epsilon_{\mu\nu} k_\mu A_\nu(k)\label{014740_1May18}.
\end{equation}
Thus, we have obtained the gauge invariant form of the fermion number anomaly with the correct normalization.%
\footnote{The coefficient $\frac{1}{4\pi}$ shows that it is the consistent anomaly.
It would be replaced with $\frac{1}{2\pi}$ if it were the covariant one.
It is interesting that the result is consistent one although our regularization is related to the covariant regularization as in Eq.\eqref{eq: rev_cov}.} 
From Eq.\eqref{014740_1May18}, it can be seen that the signature of $M$ corresponds to the chirality of the chiral fermion.

\subsection{Comments on Eq.\eqref{014740_1May18}}
We give three comments on the anomaly Eq.\eqref{014740_1May18}.
Firstly, because we have considered the Abelian case, the fermion number current takes the same form as the gauge current does. 
In non-Abelian cases, the former is defined also by Eq.\eqref{014202_1May18}.
On the other hand, the latter would be defined as 
\begin{align}
 J_\mu^a(x)&\equiv \int ds \Big[ \bar{\psi}(x,s)\gamma_\mu t^a \psi(x,s) +  \bar{\phi} (x,s)\gamma_\mu t^a \phi(x,s) \n \\
 &+ \sum_{i=1}^m\sum_{r=1}^{|C_i|} \left[ \bar{\psi}_{i,r}(x,s)\gamma_\mu t^a \psi_{i,r}(x,s) + \bar{\phi}_{i,r}(x,s)\gamma_\mu t^a \phi_{i,r}(x,s)\right] \Big].
\end{align}
Here $t^a$ is the hermitian generator in the representation of the fermion multiplet.
This would reproduce the consistent gauge anomaly in general dimensions.

Secondly, let us consider why the correct fermion number anomaly has been derived in our regularization.
Because the {\it regularized} Lagrangian Eq.(\ref{eq: rev_action_reg}) is invariant under the $U(1)$ rotation, the corresponding current $J_\mu$ seems to be conserved even at the {\it quantum level}.
However, it is not the case.
In order to see this, let the gauge field evolve slowly in the $s$-direction.
It can be realized by, for example, the gradient flow:
\begin{equation}
 \partial_s A_\mu = -\frac{\xi~\epsilon(s)}{M} \frac{\delta S[A]}{\delta A_\mu},
\end{equation}
where $\xi$ is a flow parameter and $\xi =0$ corresponds to switching off the flow.
In this situation ($\xi \neq 0$), the parity-odd part of $\Pi_{\mu\nu}^{blk}$ in Eq.\eqref{181742_8May18} does not vanish but contributes as the parity anomaly.
Thus Eq.\eqref{014740_1May18} must be replaced by 
\begin{equation}
 \braket{k_\mu J_\mu}_A =  \frac{-1}{4\pi}\frac{M}{|M|}\epsilon_{\mu\nu} k_\mu A_\nu(k,s=0) + \frac{2 M}{|M|}\int_0^\infty  ds~ k_\mu \frac{\delta S_{CS}}{\delta A_\mu(k,s)},\label{031329_1May18}
\end{equation}
where $S_{CS}$ is the three-dimensional Chern-Simons action.
The second term in Eq.\eqref{031329_1May18} can be rewritten as a surface term:
\begin{equation}
 \frac{2 M}{|M|} \int_0^\infty  ds~ k_\mu \frac{\delta S_{CS}}{\delta A_\mu(k,s)} = \left[\frac{-1}{4\pi}\frac{M}{|M|}\epsilon_{\mu\nu} k_\mu A_\nu(k,s)  \right]_{s=0}^{s=\infty}.
\end{equation}
The contribution at $s=0$ is canceled by the first term in Eq.\eqref{031329_1May18} (See Ref.\cite{Callan:1984sa}.) and thus we obtain
\begin{equation}
 \braket{k_\mu J_\mu}_A = \frac{-1}{4\pi}\frac{M}{|M|}\epsilon_{\mu\nu} k_\mu A_\nu(k,s=\infty).\label{183224_8May18}
\end{equation}
Because the gauge field becomes pure gauge at $s=\infty$, this quantity vanishes and then the current is conserved.
In our case ($\xi \to 0$), however, $A_\mu$ does not evolve: $A_\mu(k,s=\infty)=A_\mu(k)$.
Thus Eq.\eqref{183224_8May18} does not vanish and is equal to the previous result Eq.\eqref{014740_1May18}, which has been obtained by setting $\Pi_{\mu\nu}^{blk}=0$ from the beginning.
Consequently, we can understand the origin of the anomaly as follows.
In the three-dimensional region except for the boundary, the current is conserved as seen from the regularized Lagrangian.
However, this system is not closed under the condition $A_\mu(x,s)=A_\mu(x)$, and thus the current flows out to the infinity in s-direction. 
In the view point of the chiral mode on the domain-wall, this current non-conservation is seen as the anomaly.\footnote{When one considers a finite size system ($-L<s<L$) and imposes the periodic boundary condition, the current is conserved because another chiral fermion will be induced on the anti-domain wall $s=L$ that {\it absorbs} the flowing-out current . } 

Finally, let us consider another definition of the fermion number current.
Naively, in the ordinary $2n$-dimensional system consisting of one chiral fermion, the $U(1)$ current
\begin{equation}
 \bar{\psi}_L\gamma_\mu P_L\psi_L
\end{equation}
seems to be equal to the chiral $U(1)$ current 
\begin{equation}
\bar{\psi}_L\gamma_5 \gamma_\mu P_L\psi_L\label{215039_8May18}.
\end{equation}
Therefore, it appears that in our regularization we can define the fermion number current alternatively as
\begin{align}
 J_\mu^5(x)&\equiv \int ds \Big[ \bar{\psi}(x,s)\gamma_5\gamma_\mu\psi(x,s) +  \bar{\phi} (x,s)\gamma_5\gamma_\mu \phi(x,s) \n \\
 &+ \sum_{i=1}^m\sum_{r=1}^{|C_i|} \left[ \bar{\psi}_{i,r}(x,s)\gamma_5\gamma_\mu \psi_{i,r}(x,s) + \bar{\phi}_{i,r}(x,s)\gamma_5\gamma_\mu \phi_{i,r}(x,s)\right] \Big].
\end{align}
However, this is not the case, since the regularized Lagrangian is not invariant under the transformation $\psi\to e^{i\alpha \gamma_5}\psi$.
Indeed one can check that the divergence of this current differs from Eq.\eqref{014740_1May18} by both parity-odd and even terms.

\section{Partially Dimensional Regularization --Example--}\label{004700_18May18}
In this section, we propose the PDR as a novel useful regularization and demonstrate how it works.
Let us consider the 4D Yukawa theory as an example.
Here we assume that the fermion is regularized by the ordinary PV fields.
The PDR changes the spacetime dimension of the scalar field from 4 to $(4+\epsilon)$ while keeping the fermion in the 4-dimensional space.
We will show that the two parameters, the PV mass and $\epsilon$, play the roles of the ultraviolet (UV) cutoffs and regularize all the UV divergences in the theory.
In Sec.\ref{143839_21May18}, we will apply the PDR to the gauge sector in chiral gauge theory.

\subsection{Yukawa theory}\label{232655_28Apr18}
The regularized action is given by\footnote{We denote the Dirac fermion and scalar field by $\psi(x)$ and $\phi(x,x_\epsilon)$, respectively, only in this section.
The reader should not confuse them with the DW fermion and the subtracting field.}
\begin{align}
S = & \int d^{4+\epsilon}x \left[\frac{1}{2}\left(\partial_\mu\phi (x,x_\epsilon)\right)^2 + \frac{\mu^2}{2}(\phi(x,x_\epsilon))^2  + \frac{\lambda}{4!}(\phi(x,x_\epsilon))^4\right] \n \\
 &+ \int d^4x~ \bar{\psi}(x)\left[\slashed{\partial}-m + g \phi(x,0)\right] \psi(x)\label{173957_1Apr18} ,
\end{align}
where $x_\epsilon$ denotes the $\epsilon$-dimensional coordinate.
Note that we have omitted the PV fields for the fermion.
In this system, the fermion $\psi(x)$ lives on the 4-dimensional brane embedded in the $(4+\epsilon)$-dimensional space, $x_\epsilon=0$, while the scalar lives in the $(4+\epsilon)$-dimensional space.

From the Lagrangian Eq.\eqref{173957_1Apr18}, we obtain the Feynman rules as follows:
\begin{equation}
 \contraction{}{\psi}{(x)}{\bar{\psi}} \psi(x)\bar{\psi}(y) = \int \frac{ d^4 p}{(2\pi)^4} ~e^{ip\cdot(x-y)} \frac{1}{i\slashed{p}-m} ,\label{210644_29Apr18}
\end{equation}
\begin{equation}
 \contraction{}{\phi}{(x,x_\epsilon)}{\phi} \phi(x,x_\epsilon) \phi(y,y_\epsilon) = \int \frac{d^{4+\epsilon} k}{(2\pi)^{4+\epsilon}}~ e^{ik\cdot(x-y)+i k_\epsilon (x_\epsilon-y_\epsilon)} \frac{1}{k^2 + (k_\epsilon)^2 + \mu^2} ,\label{210656_29Apr18}
\end{equation}
%
\begin{center}
 \begin{tabular}{c}
      \begin{minipage}{0.25\hsize}
        \begin{center}
          \includegraphics[ width=1.0\textwidth]{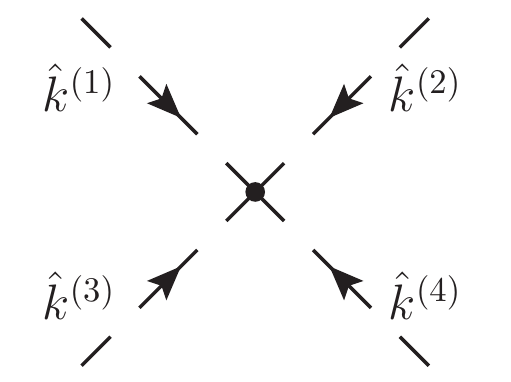}
        \end{center}
      \end{minipage}

      \begin{minipage}{0.45\hsize}
      \begin{equation}
       = - \lambda ~\delta^{(4+\epsilon)}(\hat{k}^{(1)}+\hat{k}^{(2)}+\hat{k}^{(3)}+\hat{k}^{(4)}),\label{210704_29Apr18}
      \end{equation}
      \end{minipage}
 \end{tabular}
\end{center}
%
\begin{center}
 \begin{tabular}{c}
      \begin{minipage}{0.25\hsize}
        \begin{center}
          \includegraphics[ width=1.0\textwidth]{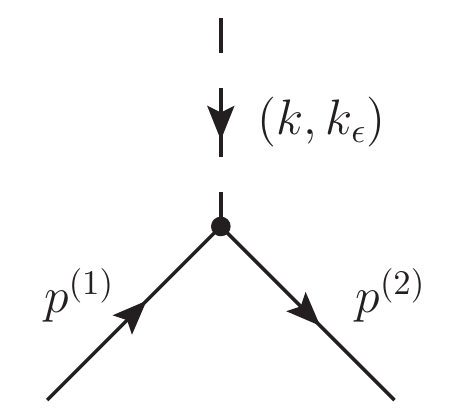}
        \end{center}
      \end{minipage}

      \begin{minipage}{0.45\hsize}
       \begin{equation}
       = - g~ \delta^{(4)}(p^{(1)} -p^{(2)} + k) .\label{232354_9Apr18}
       \end{equation}
      \end{minipage}
 \end{tabular}
\end{center}
 %
Here $\hat{k}$ denotes the $(4+\epsilon)$-dimensional momentum, i.e. $\hat{k}=(k,k_\epsilon)$, and we have written explicitly the $\delta$-functions enforcing the conservation of momentum.
Note that at the Yukawa vertex, Eq. \eqref{232354_9Apr18}, the $\delta$-function is only for the 4-dimensions, and hence the $\epsilon$-dimensional momentum $k_\epsilon$ of the scalar $\phi(\hat{k})$ is not conserved.

\begin{figure}[tbp]
  \begin{center}
  \includegraphics[width=0.35\textwidth]{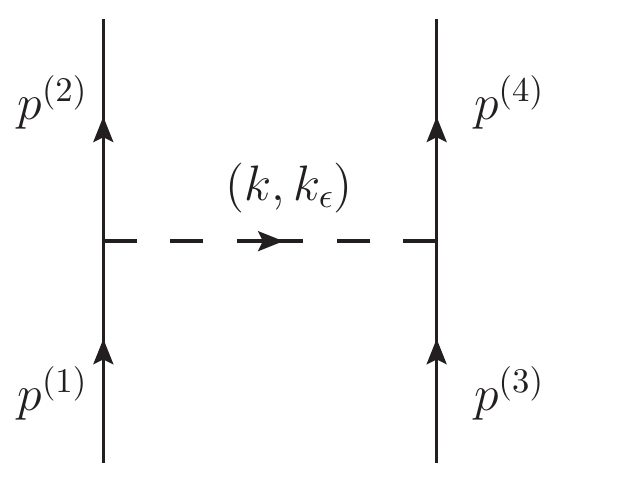}
  \caption{A diagram including a scalar internal line with the Yukwa vertices at the ends}
\label{183420_5Apr18}
  \end{center}
 \end{figure}

 Under the above rules, we have to pay a special attention when dealing with a diagram including an internal scalar line with the Yukawa vertices at the ends.
 As an example, let us consider a diagram shown in Fig.\ref{183420_5Apr18}.
The $\epsilon$-dimensional momentum $ k_\epsilon$ flows along the scalar line but does not along the fermion ones.
In the Fourier space, the internal line gives
\begin{equation}
 \int \frac{d^\epsilon k_\epsilon}{(2\pi)^\epsilon}~\frac{1}{k^2+(k_\epsilon)^2 +\mu^2}~.
\end{equation}
The integral over $k_\epsilon$ appears because it is not constrained by any $\delta$-functions.
Carrying out the integral, we obtain
\begin{equation}
  \frac{\Gamma(1-\epsilon/2)}{(4\pi)^{\epsilon/2}} \frac{1}{[k^2 + \mu^2]^{1-\epsilon/2}} .\label{222356_28Apr18}
\end{equation}
The exponent ($1-\epsilon/2$) in the denominator plays a role of regularization for loop diagrams, as will be seen in the next subsection.

\subsection{One-loop calculation}
\begin{figure}[tbp]
  \begin{center}
  \includegraphics[width=0.5\textwidth]{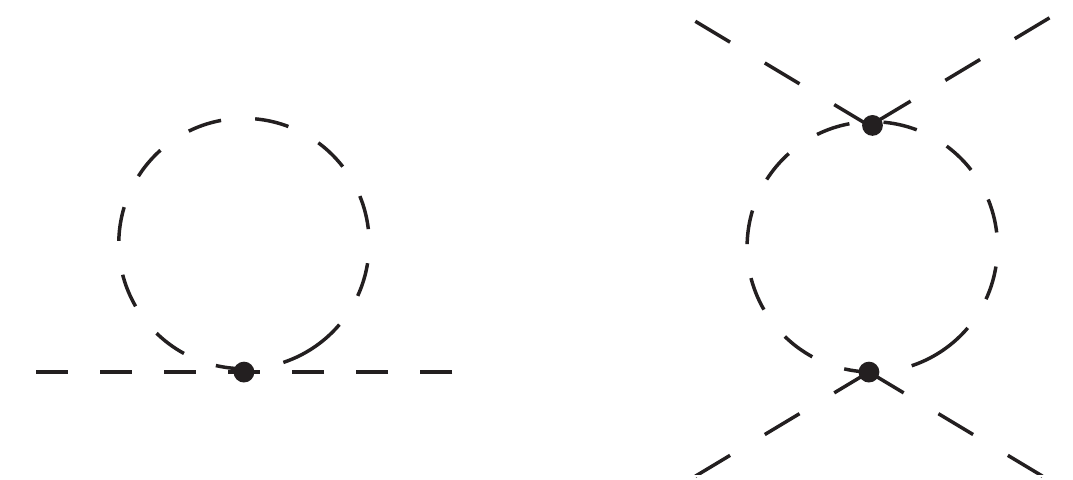}
  \caption{One-loop diagrams consisting of only the scalar lines}
\label{225530_29Apr18}
  \end{center}
  \end{figure}

\begin{figure}[tbp]
  \begin{center}
  \includegraphics[width=0.45\textwidth]{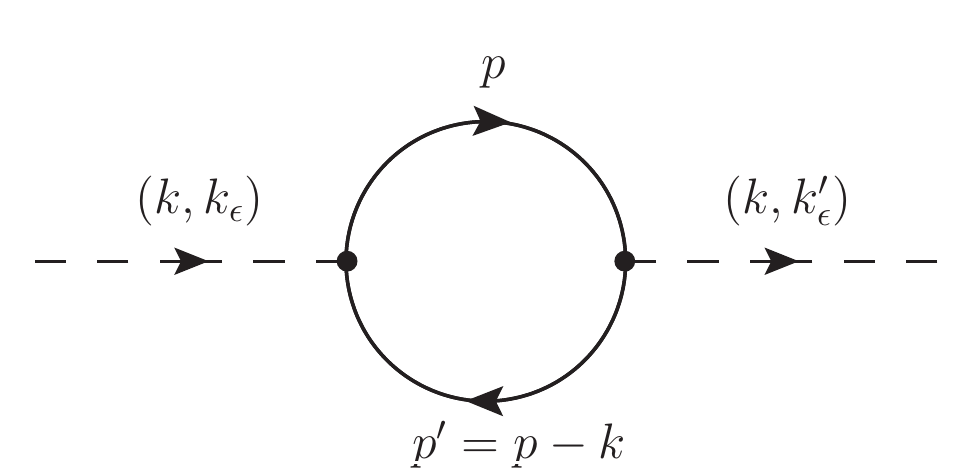}
  \caption{A fermion loop diagram}
\label{225636_29Apr18}
  \end{center}
\end{figure}

\begin{figure}[tbp]
\begin{center}
 \begin{tabular}{c}
      \begin{minipage}{0.45\hsize}
        \begin{center}
          \includegraphics[ width=0.9\textwidth]{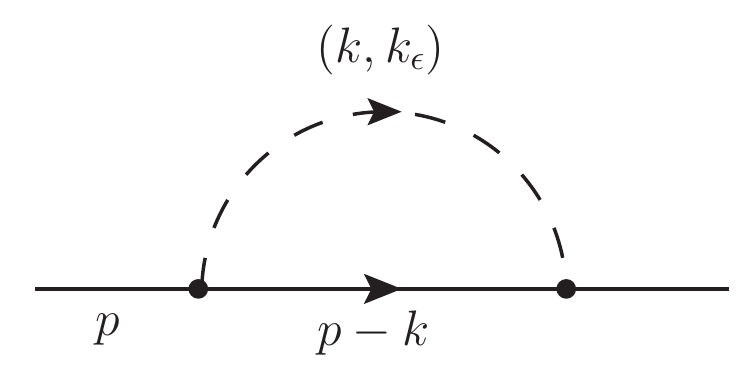}
         \\ (a) Fermion self-energy $\Sigma(p)$
        \end{center}
      \end{minipage}

      \begin{minipage}{0.45\hsize}
        \begin{center}
          \includegraphics[ width=0.7\textwidth]{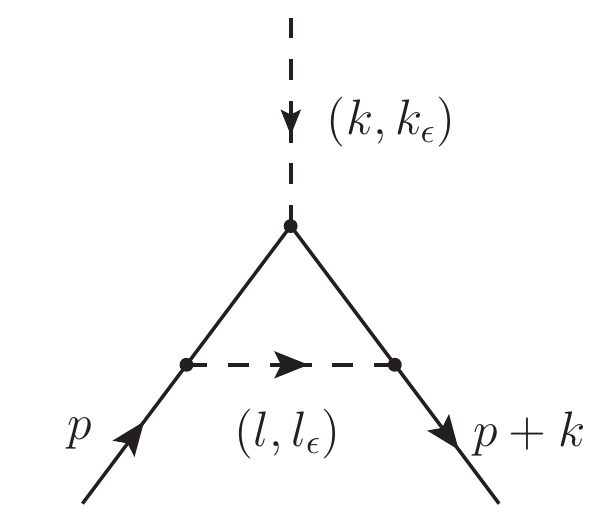}
         \\ (b) Yukawa vertex correction $\Gamma(p,k)$
        \end{center}
      \end{minipage}
 \end{tabular}

 \caption{One-loop diagrams including both of the scalar and fermion internal lines}
  \label{203402_5Apr18}
\end{center}
\end{figure}

We show that how our method can regularize the divergences in the one-loop diagrams.
 Clearly, diagrams consisting of only the scalar lines, which are shown in Fig.\ref{225530_29Apr18}, are regularized in the same manner as in the ordinary dimensional regularization.
 On the other hand, a fermion loop, shown in Fig.\ref{225636_29Apr18}, is regularized by the PV fields 
 \footnote{Note that the $\epsilon$-dimensional momentum in the external $\phi$ lines is not conserved, that is $k_\epsilon\neq k_\epsilon'$, 
because the Yukawa vertex conserves momentum only in the 4-dimensions.
 However, this does not affect the unitarity as the cutoff is removed ($\epsilon\to 0$).}.
Therefore, what we are interested in is ones involving both the scalar and fermion internal lines. (See Fig.\ref{203402_5Apr18}.)

  %

Let us consider the fermion self-energy diagram Fig.\ref{203402_5Apr18}(a), which we denote by $\Sigma(p)$.
Using Eq.\eqref{222356_28Apr18}, we have
\begin{align}
 \Sigma(p)&=g^2~\int \frac{d^4k}{(2\pi)^4}~\frac{1}{i(\slashed{p}-\s{k})-m}  ~\frac{\Gamma(1-\epsilon/2)}{(4\pi)^{\epsilon/2}}\frac{1}{[k^2+\mu^2]^{1-\epsilon/2}} \\
 & = -g^2 \frac{\Gamma(1-\epsilon/2)}{(4\pi)^{\epsilon/2}} \int \frac{d^4k}{(2\pi)^4}\frac{i(\s{p}-\s{k}) +m }{(p-k)^2+m^2} ~\frac{1}{[k^2+\mu^2]^{1-\epsilon/2}}.
\end{align}
Its integration over $k$ leads to the following expression:
\begin{equation}
 \Sigma(p) = -g^2 \frac{\Gamma(-\epsilon/2)}{(4\pi)^{2+\epsilon/2}}\int_0^1 dx~x^{-\epsilon/2} \left[i(1-x)\s{p} +m\right] (\Delta')^{\epsilon/2} \label{180935_29Apr18}
\end{equation}
with
\begin{equation}
 \Delta' \equiv x(1-x)p^2 + x m^2 + (1-x)\mu^2.
\end{equation}
Here $x$ is the Feynman parameter.
This expression Eq.\eqref{180935_29Apr18} is finite as long as $0<\epsilon<2$, and thus the UV divergence is regularized in our model. 
Furthermore, one can check that its difference from the result in the ordinary dimensional scheme is finite constants.
%

Next, we investigate the other diagram (Fig.\ref{203402_5Apr18}(b)), which is the one-loop vertex correction $\Gamma(p,k)$.
Using Eqs.\eqref{210644_29Apr18}--\eqref{232354_9Apr18} and \eqref{222356_28Apr18}, we have
\begin{align}
 \Gamma(p,k) = g^3 \int \frac{d^4 l}{(2\pi)^4} ~\frac{1}{i(\slashed{p}-\slashed{l}+\slashed{k})-m}~\frac{1}{i(\slashed{p}-\slashed{l})-m}~  \frac{\Gamma(1-\epsilon/2)}{(4\pi)^{\epsilon/2}} \frac{1}{[l^2 + \mu^2]^{1-\epsilon/2}}\label{161302_3Apr18}.
\end{align}
After some calculation, we obtain the following result:
\begin{align}
 \Gamma(p,k)= \frac{g^3}{16\pi^2} \int_{xyz}\left(\frac{-2}{(4z\pi)^{\epsilon/2}}\Gamma(-\epsilon/2)(\Delta'')^{\epsilon/2} + \frac{\left[i(z\s{p}+(1-x)\s{k})+m\right]\left[i(z\s{p}-x\s{k})+m\right]}{\Delta''} \right)
\end{align}
with
\begin{equation}
\int_{xyz} \equiv \int_0^1  dx \int_0^1 dy \int_0^1 dz ~\delta(x+y+z-1) , 
\end{equation}
\begin{equation}
 \Delta'' \equiv x(1-x)(p+k)^2 +y(1-y)p^2 - 2xy (p+k)\cdot p +(x+y)m^2 + z \mu^2 .
\end{equation}
Also in this case, the UV divergence is controlled thanks to $\epsilon$.

Thus, the combination of the PDR and the PV regularization can regularize all the one-loop divergences.
It is obvious that this is also true about higher-order loop diagrams. 
Note that the PDR is essentially as a sort of the analytic regularization \cite{Sp1,We,Speer:1972wz}, since the integration over $k_\varepsilon$ gives $k$ of the fractional power. 
%

\section{PDR for Chiral Gauge Theory}\label{143839_21May18}
In Sec.\ref{sec: DW} and \ref{sec: chiral_anomaly}, we have regularized the fermion sector only.
For practical calculations, however, we must also regularize the gauge sector.
In this section, we apply the PDR to the gauge sector while keeping the DW fermion on the 5-dimensional space.
After that, we will show that all the one-loop UV divergences are regularized by the combination of the PDR and the PV regularization, and that they can be renormalized by local counter terms.
For simplicity, we consider the case where the resulting physical theory is 4-dimensional.

\subsection{The regularized model}
\begin{figure}[tbp]
  \begin{center}
  \includegraphics[width=0.4\textwidth]{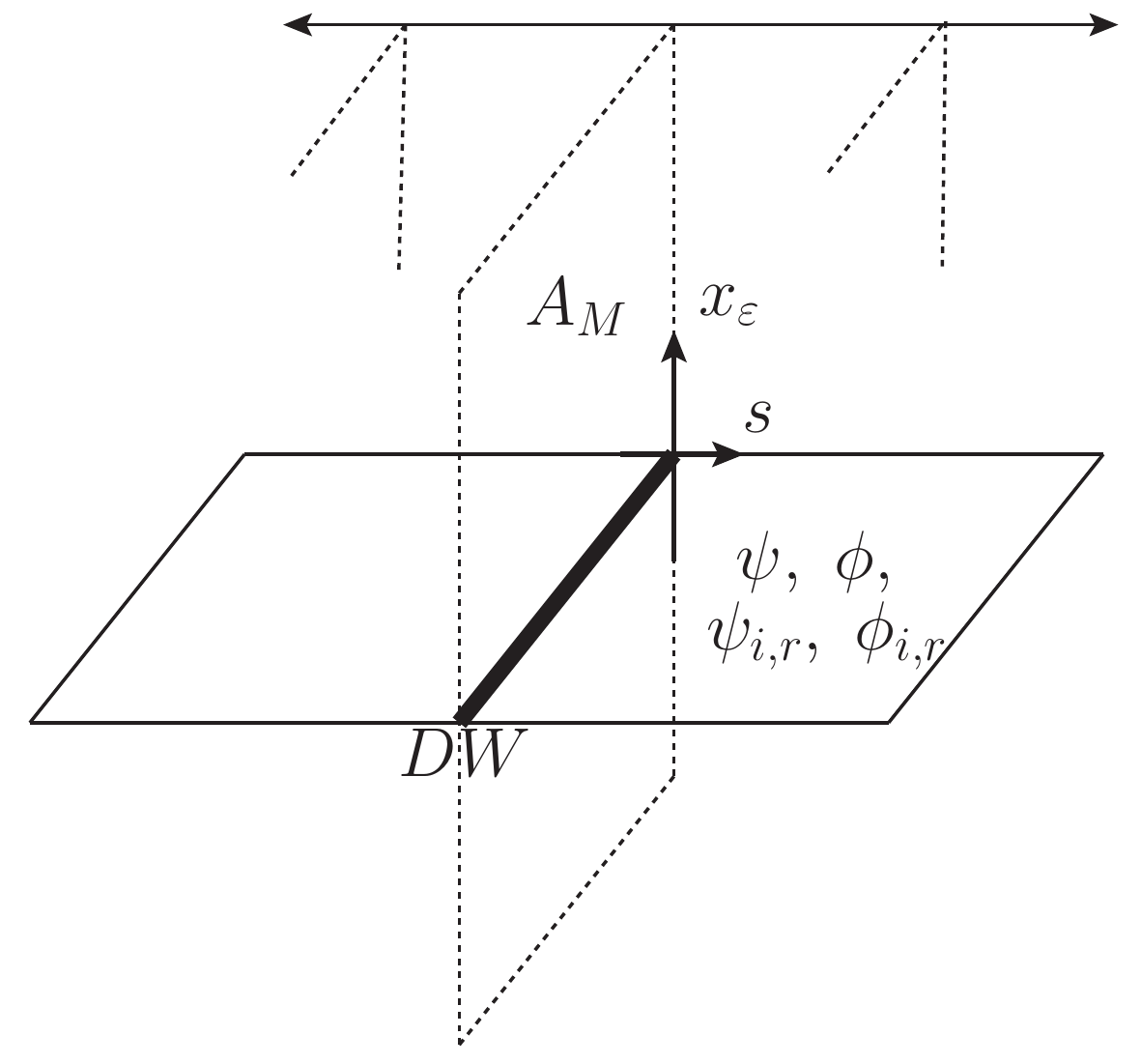}
  \caption{The set up of our model Eq.\eqref{185405_1Jun18} in the coordinate space. $\psi$'s and $\phi$'s live on the $(4+1)$-dimensional brane $x_\varepsilon=0$ (solid brane). Their massless modes are localized on the 4-dimensional domain-wall (the thick line). On the other hand, $A_M$ is defined as the gauge field on the $(4+\varepsilon)$ dimensional brane $s=0$ (the dotted brane), and it is duplicated along the $s$-direction.}
\label{fig: setup}
  \end{center}
 \end{figure}

Firstly, we replace the 4-dimensional gauge field $A_\mu(x)$ in Eq.\eqref{eq: rev_action_reg} with the $(4+\epsilon)$-dimensional one $A_M(x,x_\epsilon)$.
Here $x$ and $x_\epsilon$ are the coordinates of the 4-dimensional and $\epsilon$-dimensional space, respectively, and the lower index $M$ runs from $1$ to $4+\epsilon$.
Next, we extend the gauge field in the $s$-direction, and thus the gauge field lives in the $(4+\epsilon+1)$-dimensional space.
Note that it is independent of $s$, that is $A_M(x,x_\epsilon,s)=A_M(x,x_\epsilon)$.
Furthermore, we set the $(4+\epsilon+1)$th component of the field to 0: $A_{4+\epsilon+1}=0$.
On the other hand, the DW fermion (and all the regulator fields) live on the 5-dimensional brane $x_\epsilon=0$ embedded in the $(4+\epsilon+1)$-dimensional space.
Therefore, the regularized action of this system is given by the following equation:
\begin{align}
 S_{reg} &= \int d^4x ds ~\bar{\psi}\left[\slashed{D}+\gamma_5 \partial_s -M\epsilon(s)\right]\psi + \int d^4x ds ~\bar{\phi}\left[\slashed{D}+\gamma_5 \partial_s +M\right]\phi \n \\
 & + \sum_{i=1}^{m}\sum_{r=1}^{|C_i|}\int d^4x ds ~\bar{\psi}_{i,r}\left[\slashed{D}+\gamma_5 \partial_s -M\epsilon(s)\right]\psi_{i,r} \n \\
 & + \sum_{i=1}^{m}\sum_{r=1}^{|C_i|} \int d^4x ds ~\bar{\phi}_{i,r}\left[\slashed{D}+\gamma_5 \partial_s +M\right]\phi_{i,r} \n \\
 & +\frac{1}{4g^2} \int d^{4+\epsilon} x~ \mathrm{tr}(F_{MN}(x,x_\epsilon))^2,\label{185405_1Jun18}
\end{align}
with
\begin{equation}
 D_\mu = \partial_\mu-iA_\mu(x,0) \hspace{2em}(\mu=1,\cdots,4) .
\end{equation}
Fig.\ref{fig: setup} shows this model in the coordinate space.

In terms of $G_\psi$ and $G_\phi$ that have been defined by Eqs.\eqref{154523_24May18}-\eqref{154629_24May18}, the Feynman rules are expressed as follows:
\begin{equation}
 \contraction{}{\psi}{(x,s)}{\bar{\psi}} \psi(x,s)\bar{\psi}(y,s') = \int \frac{d^4p}{(2\pi)^4}~e^{ip\cdot (x-y)}~G_\psi (p,s,s')
\end{equation}
\begin{equation}
 \contraction{}{\phi}{(x,s)}{\bar{\phi}} \phi(x,s)\bar{\phi}(y,s') =  \int \frac{d^4p}{(2\pi)^4}~e^{ip\cdot (x-y)} ~G_\phi (p,s,s')
\end{equation}
\begin{equation}
 \contraction{}{A_\mu^a}{(x,x_\epsilon)}{A_\nu^b} A_\mu^a(x,x_\epsilon) A_\nu^b(y,y_\epsilon) = \int \frac{d^{4+\epsilon} }{(2\pi)^{4+\epsilon}}~ e^{ik\cdot(x-y)+ik_\epsilon\cdot(x_\epsilon-y_\epsilon)}~ \frac{\delta^{ab}\delta_{\mu\nu}}{k^2 + k_\epsilon^2} \h{2em} \mr{(Feynman~ gauge)}
\end{equation}
\begin{center}
 \begin{tabular}{c}
      \begin{minipage}{0.25\hsize}
        \begin{center}
          \includegraphics[ width=1.0\textwidth]{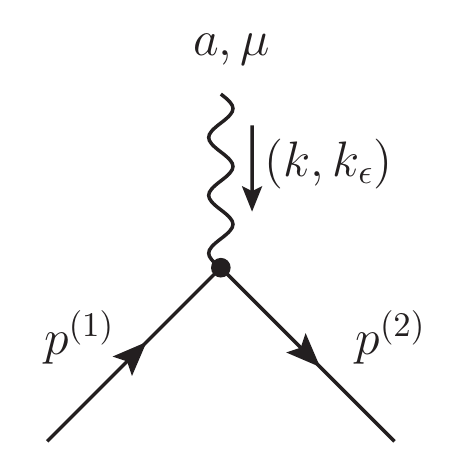}
        \end{center}
      \end{minipage}

      \begin{minipage}{0.45\hsize}
      \begin{equation}
       =  - ig t^a \gamma_\mu \delta^{(4)}(p^{(1)} +p^{(2)} + k),
      \end{equation}
      \end{minipage}
 \end{tabular}
\end{center}
%
Here we have written explicitly the delta-functions enforcing momentum conservation. 
We also have the gauge boson's three- and four-point vertices which are the same form as in the ordinary dimensional regularization. 
Note that the $\epsilon$-dimensional momentum is not conserved at the gauge-fermion coupling.
This situation is similar to the the Yukawa theory in the previous section. 

Before calculating the one-loop diagrams, let us consider a diagram containing an internal line of $A_\mu$ with the gauge-fermion vertices at the ends.
Similarly to the previous section, the internal line gives 
\begin{align}
 \int \frac{d^\epsilon k_\epsilon}{(2\pi)^\epsilon} \frac{\delta_{\mu\nu} \delta^{ab}}{k^2+(k_\epsilon)^2} = \frac{\Gamma(1-\epsilon/2)}{(4\pi)^{\epsilon/2}} \frac{\delta_{\mu\nu} \delta^{ab}}{(k^2)^{1-\epsilon/2}}\label{123750_1May18}.
\end{align}
The exponent $(1-\epsilon/2)$ plays a role of regularization as in Sec.\ref{004700_18May18}.

  \subsection{One-loop calculation}
\begin{figure}[tbp]
\begin{center}
 \begin{tabular}{c}
      \begin{minipage}{0.35\hsize}
        \begin{center}
          \includegraphics[ width=1.0\textwidth]{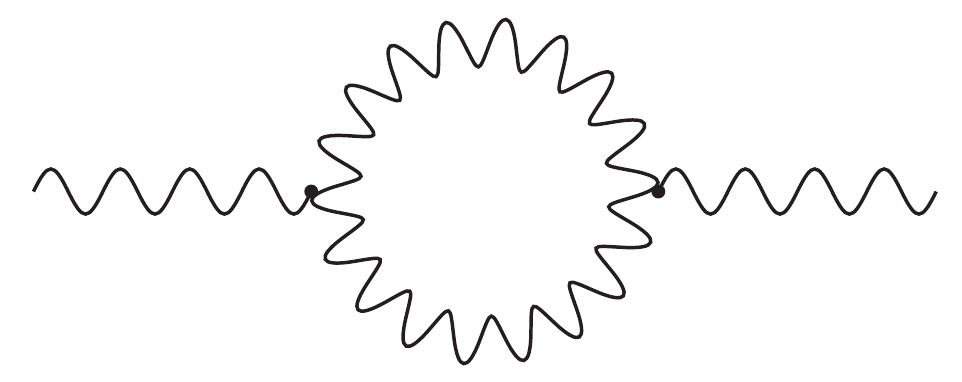}
        \end{center}
      \end{minipage}

      \begin{minipage}{0.35\hsize}
        \begin{center}
          \includegraphics[ width=0.6\textwidth]{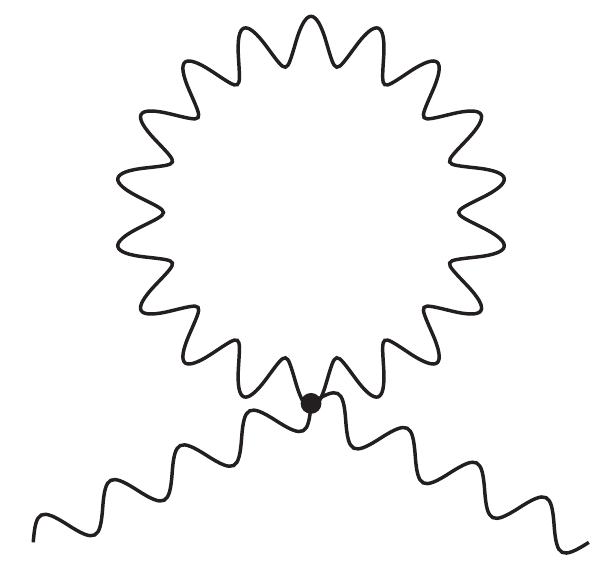}
        \end{center}
      \end{minipage}
 \end{tabular}
    \caption{The gauge boson one-loop diagrams}
\label{054218_9Apr18}
\end{center}
 \end{figure}

\begin{figure}[tbp]
  \begin{center}
  \includegraphics[width=0.45\textwidth]{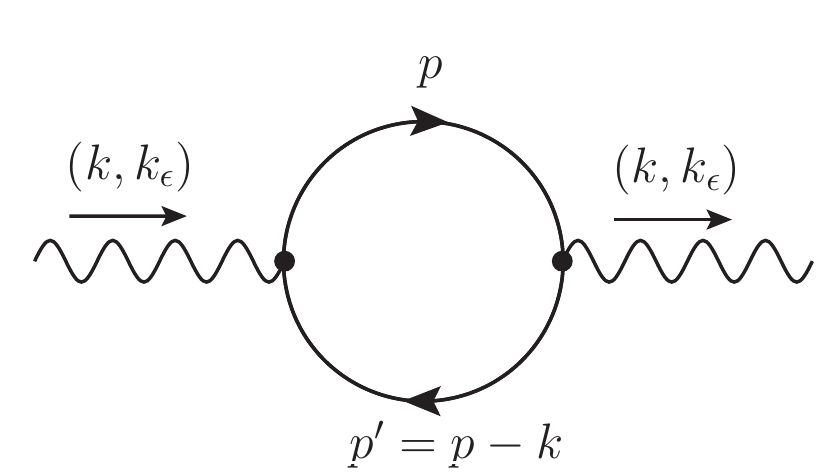}
  \caption{The vacuum polarization diagram}
\label{051529_9Apr18}
  \end{center}
\end{figure}

Firstly, we consider the gauge boson loop diagrams and the fermion loop diagram (See Fig.\ref{054218_9Apr18} and Fig.\ref{051529_9Apr18}).
It is obvious that the UV divergences of the former are regularized and renormalized to the $(4+\epsilon)$-dimensional gauge kinetic term as in the ordinary dimensional scheme.
The latter are regularized by the subtracting field and the PV pairs as in Sec.\ref{sec: DW}.
Here note that the $\epsilon$-dimensional momenta in the external gauge boson lines are not conserved: 
$k_\epsilon\neq k_\epsilon'$ because only 4-dimensional momentum is conserved at the fremion-fermion-gauge boson vertex. 
Therefore the UV divergence is renormalized by the {\it 4-dimensional} local counter term 
\begin{equation}
 \delta_g \int d^4x~ \mr{tr}(F_{\mu\nu})^2 .
\end{equation}
This term has the $(4+\varepsilon)$-dimensional gauge invariance even when $\varepsilon$ is finite, and does not cause any problem in the limit $\varepsilon\to0$. 
 %
%
%

\begin{figure}[tbp]
  \begin{center}
  \includegraphics[width=0.5\textwidth]{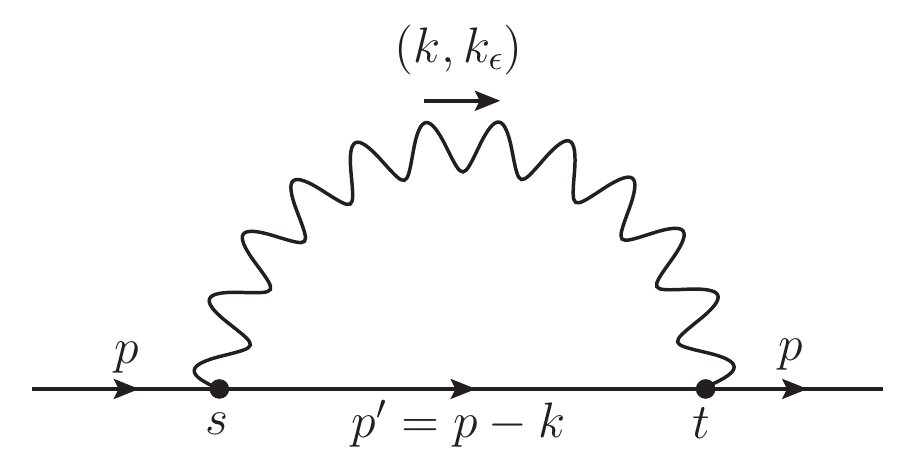}
  \caption{The fermion self-energy diagram}
\label{051405_9Apr18}
  \end{center}
 \end{figure}

Next, we consider the DW fermion's self-energy diagram (Fig.\ref{051405_9Apr18}).
This diagram induces radiative corrections to the massless and the bulk propagators because the external lines contain the both modes.
Obviously, it is not easy to investigate what counter terms are necessary.
Therefore, we adopt the following strategy.
We start with analyzing the self-energy in the region far from the domain wall, i.e. $s \gg M^{-1}$, where the massless mode is suppressed and only the bulk modes remain in the diagram.
Furthermore, the argument of renormalization is parallel to the ordinary {\it 4-dimensional} Dirac fermion because the gauge field does not vary in the $s$-direction.
After that, we consider the case that $M$ is large compared with the external momentum.
This limit is equivalent to focusing on the correction to the massless propagator.
As will see below, the counter terms that are necessary for the bulk modes in the region $s \gg M^{-1}$ are sufficient to renormalize the massless mode.

%

For $s \gg M^{-1}$, we can evaluate the diagram in the 5-dimensional Fourier space because the propagator of the bulk modes are translation invariant.
Noting that the momentum of the gauge field does not have the 5-th component, we have
\begin{align}
 &\int \frac{d^4 k}{(2\pi)^4}(i\gamma_\mu t^a)\frac{1}{i(\s{p}-\s{k}) + i\gamma_5 p_5 -M}(i\gamma_\nu t^b) \frac{\Gamma(1-\epsilon/2)}{(4\pi)^{\epsilon/2}}\frac{\delta_{\mu\nu}\delta^{ab}}{(k^2)^{1-\epsilon/2}} \\
&=(ig)^2 t^a t^a \frac{\Gamma(1-\epsilon/2)}{(4\pi)^{\epsilon/2}} \int \frac{d^4 k}{(2\pi)^4} ~ \frac{2i(\s{p}-\s{k})-4M'}{(p-k)^2 + |M'|^2 } ~\frac{1}{(k^2)^{1-\epsilon/2}}\label{180934_31May18} 
\end{align}
with
\begin{equation}
 M' \equiv -i\gamma_5p_5 +M.
\end{equation}
Here we have denoted the 5-th component of momentum of the fermion by $p_5$ .
Note that the expression \eqref{180934_31May18} seems to be the self-energy in the 4-dimensional theory with the effective mass $M'$.
Therefore it is obvious that the UV divergences can be renormalized by the ordinary procedure in the 4-dimensions.
In order to obtain the counter terms that eliminate the UV divergences, we expand Eq.\eqref{180934_31May18} by $p$:
\begin{align}
 &(ig)^2 t^a t^a \frac{\Gamma(1-\epsilon/2)}{(4\pi)^{\epsilon/2}} \int \frac{d^4 k}{(2\pi)^4} ~\frac{2}{(k^2)^{1-\epsilon/2}} \times \n \\
& \h{2em} \left[\frac{-i\s{k}-2M'}{k^2 + |M'|^2} +  \frac{i\s{p}}{k^2 +|M'|^2 }+\frac{-2i\s{k}-4M'}{(k^2 + |M'|^2)^2}~p\cdot k + \mathcal{O}(p^2) \right].
\end{align}
We concentrate on the UV divergent terms and obtain 
\begin{align}
(ig)^2 t^a t^a \frac{\Gamma(1-\epsilon/2)}{(4\pi)^{\epsilon/2}} \int \frac{d^4 k}{(2\pi)^4} ~\frac{1}{(k^2)^{1-\epsilon/2}}  \left[\frac{-4M'}{k^2 + |M'|^2} +  i\s{p} \frac{k^2}{(k^2 + |M'|^2 )^2}  \right] .
\end{align}
After some calculation, we obtain the following expression:
\begin{align}
& (ig)^2 t^a t^a  ~\Gamma(-\epsilon/2) \left[\frac{-M'}{4\pi^2} \int_0^1 dx \left(\frac{x|M'|^2}{4\pi(1-x)}\right)^{\epsilon/2}+ \frac{i\s{p}}{8\pi^2} \int_0^1 dx ~ x \left(\frac{x |M'|^2}{4\pi(1-x)}\right)^{\epsilon/2}\right] \\
&\to (ig)^2 t^a t^a ~ \Gamma(-\epsilon/2) \left[\frac{-M'}{4\pi^2}+ \frac{i\s{p}}{16\pi^2} \right] \h{1em}(\mr{as}~~\epsilon\to 0).
\end{align}
It can be seen that the UV divergences are regularized by $\epsilon$ as in Sec.\ref{004700_18May18}.
Consequently, the necessary counter terms are found to be
 \footnote{The difference between the coefficients of $\s{\partial}$ and $\gamma_5\partial_s$ reflects the lack of the 5-dimensional Lorentz invariance.
On the other hand, $\gamma_5 \partial_s$ and $M$ share the same coefficient because they are regarded as the effective mass in the 4-dimensions.
}
\begin{equation}
 \int d^4x \int_{s>0}ds ~\bar{\psi}\left[\delta_Z \s{\partial} - \delta_M M'\right] \psi,\label{085305_1Jun18}
\end{equation}
where
\begin{equation}
 \delta_Z =  \frac{ (ig)^2 t^at^a}{8\pi^2} \frac{1}{\epsilon},\label{090811_1Jun18}
\end{equation}
\begin{equation}
\delta_M =  \frac{(ig)^2 t^at^a}{2\pi^2} \frac{1}{\epsilon} .\label{170145_1Jun18}
\end{equation}
Note that Eqs.\eqref{090811_1Jun18} and \eqref{170145_1Jun18} are mass-independent renormalizations.
In particular, the mass is renormalized multiplicatively.
Therefore, we expect that Eqs.\eqref{090811_1Jun18} and \eqref{170145_1Jun18} hold even if the mass $M$ depends on $s$:
\begin{equation}
 \int d^4x \int ds ~\bar{\psi}\left[\delta_Z \s{\partial} + \delta_M\left(\gamma_5 \partial_s -M\epsilon(s)\right)\right] \psi \label{172017_1Jun18}.
\end{equation}
In other words, these counter terms would also be sufficient to renormalize the UV divergences of the massless mode.

 \begin{figure}[tbp]
  \begin{center}
  \includegraphics[width=0.6\textwidth]{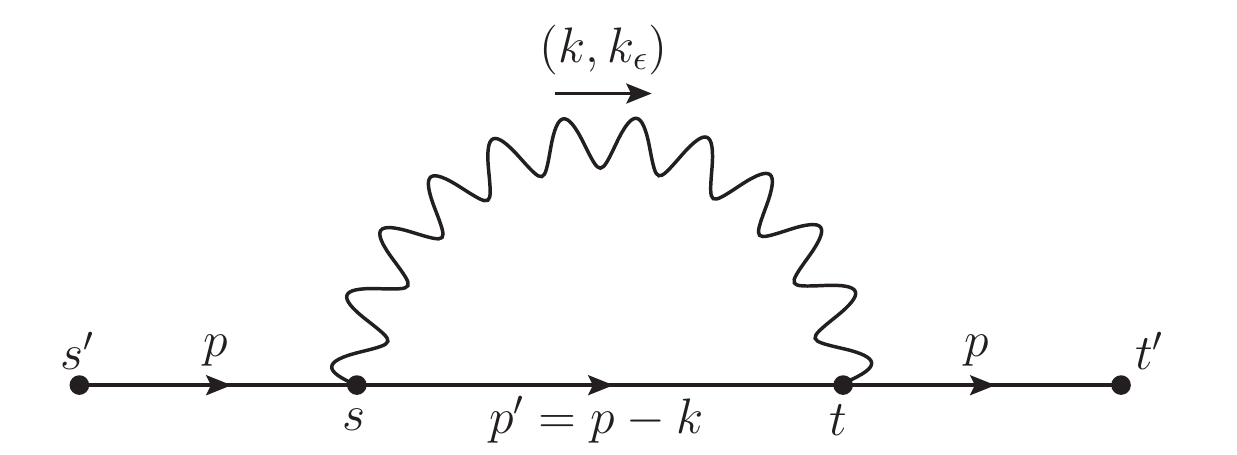}
  \caption{The fermion self-energy diagram with external legs}
  \label{173938_1Jun18}

  \end{center}
 \end{figure}
%

Let us check this point.
In order to focus on the self-energy of the massless mode, we attach the external legs to it (See Fig.\ref{173938_1Jun18}) and take the limit $M\gg p$.
In this situation, the external propagators $G_\psi(p,s,s')$ and $G_\psi(p,t',t)$ reduce to 
%
\begin{equation}
 G_\psi(p,t',t)\to -\frac{i\slashed{p}M}{p^2}P_{R} ~ e^{-(|t|+|t'|)|M|}\label{232241_23May18}
\end{equation}
\begin{align}
 G_\psi(p,s,s')\to -\frac{i\slashed{p}M}{p^2}P_{R} ~ e^{-(|s|+|s'|)|M|} = -P_L \frac{i\slashed{p}M}{p^2} ~ e^{-(|s|+|s'|)|M|} \label{173521_20May18} ,
\end{align}
where we have assumed $M>0$.
(For $M<0$, the chirality projection should be replaced with $P_L$.)
Eqs.\eqref{232241_23May18} and \eqref{173521_20May18} are nothing but the propagators of the chiral fermion localized on the domain wall.
Because the self-energy is put between $P_{R}$ and $P_{L}$, we can drop the terms except for those proportional to $\s{p}-\s{k}$.
This means that no additional mass counter terms for the massless mode are needed, and hence we can focus on the wave function renomalization.
%
%
%
We evaluate the self-energy part keeping the 5-th coordinate in the real space as follows:
\begin{align}
\frac{\Gamma(1-\epsilon/2)}{(4\pi)^{\epsilon/2}} ~\int \frac{d^4k}{(2\pi)^4}\int ds dt ~(ig)^2 t^at^a ~\frac{1}{(k^2)^{1-\epsilon/2}} \gamma_\mu G_\psi(p-k,t,s) \gamma_\mu  ~ e^{-(|s|+|t|)M}\label{035407_9Apr18} .
\end{align}
%
Substituting Eqs.\eqref{154523_24May18}-\eqref{154629_24May18} into the internal propagator $G_\psi(p-k,t,s)$\footnote{Note that one cannot take the same limit as Eqs.\eqref{232241_23May18} and \eqref{173521_20May18} for the internal propagator because the momentum in the internal line should be integrated.} and using the symmetry under the reflection $s \leftrightarrow -s$, $t\leftrightarrow -t$, we obtain the following expression:
\begin{align}
& (ig)^2 t^at^a \frac{2\Gamma(1-\epsilon/2)}{(4\pi)^{\epsilon/2}} ~\int \frac{d^4k}{(2\pi)^4} ~\frac{1}{(k^2)^{1-\epsilon/2}} \n \\
 &\h{3em} \left[ \int _0^\infty ds\int_0^\infty dt ~\frac{i\s{p}'M(\sqrt{p'^2+M^2}+M)}{p'^2 (\sqrt{p'^2+M^2})}~e^{-(s+t)(\sqrt{p'^2+M^2}+M)}\right. \n \\
 & \h{3em}+\int_0^\infty ds \int_s ^\infty dt ~\frac{i\s{p}'}{\sqrt{p'^2+M^2}}~e^{-t(\sqrt{p'^2+M^2}+M)-s(M-\sqrt{p'^2+M^2})} \n \\
 &  \h{3em}+\int_0^\infty dt \int_t ^\infty ds ~\frac{i\s{p}'}{\sqrt{p'^2+M^2}}~e^{-s(\sqrt{p'^2+M^2}+M)-t(M-\sqrt{p'^2+M^2})} \n \\
 & \h{3em}\left. + \int_0^\infty ds \int _{-\infty}^0 dt \frac{i\s{p}'(\sqrt{p'^2+M^2}-M)}{p'^2} ~e^{-(s-t)(\sqrt{p'^2+M^2}+M)} \right].
\end{align}
Here, the first and fourth terms come from the massless mode in $G_\psi(p-k,t,s)$ while the second and third ones from the bulk modes.
Note that the latter also contribute to the wave function renormalization of the massless mode.
Carrying out the integrals over $s$ and $t$, we obtain a simple result:
\begin{align}
 & \frac{-2 g^2}{M}t^at^a \frac{\Gamma(1-\epsilon/2)}{(4\pi)^{\epsilon/2}} \int\frac{d^4k}{(2\pi)^4} ~\frac{1}{(k^2)^{1-\epsilon/2}} \frac{i\s{p}'}{p'^2} \n \\
 &=\frac{-g^2}{ 8\pi^2 M}t^at^a ~i\s{p} \int _0^1 dx ~ (1-x)\left(\frac{(1-x) p^2}{4\pi }\right)^{\epsilon/2}~\Gamma(-\epsilon/2)\label{034834_9Apr18} .
\end{align}
The divergent term in Eq.\eqref{034834_9Apr18}, 
\begin{equation}
 -\frac{(ig)^2t^at^a}{8\pi^2 M}i\s{p}\frac{1}{\epsilon},\label{172704_1Jun18}
\end{equation}
can be renormalized by the counter term Eq.\eqref{172017_1Jun18}.
Indeed, putting the counter term between the external legs and integrating over $s$, we obtain
\begin{align}
 \int ds ~  \delta_Z ~i\s{p}~e^{-2|s|M} = \frac{(ig)^2t^at^a}{8\pi^2 M}i\s{p}\frac{1}{\epsilon},
\end{align}
which cancels the UV divergent term Eq.\eqref{172704_1Jun18}.

Thus, we conclude that the UV divergences in the DW fermion self-energy can be renormalized by the counter term Eq.\eqref{172017_1Jun18}.
It is easy to check that the vertex correction diagram would be renormalized in the same manner:
\begin{equation}
 \int d^4x \int ds ~\delta_Z' \bar{\psi}(x,s)\s{A}(x,x_\epsilon=0) \psi(x,s).
\end{equation}
%

\section{Summary}
\label{sec: summary}
\quad
In this paper we have proposed a regularization that incorporates with the DW fermion. 
It regulates fermion loops by introducing the PV pairs, and regulates the other loops by applying the PDR. 
By considering three-dimensional theory, we have explicitly shown that the regularization works both in parity-even and odd parts, 
and obtained the correct Abelian anomaly in the two dimensional chiral gauge theory. 
It is a good regularization of chiral gauge theory on the DW since it produces no fake anomaly. 
A significant point of this formulation is that it is given at the Lagrangian level, that is, the theory is well-defined from the starting point,  
and we need not introduce counter terms to recover the gauge symmetry. 
Furthermore, our regularization might provide a rather easy method for the explicit calculation of loop corrections. 
We also have investigated the renormalization of the DW fermion. 
There we have checked that once we add the counter terms to the full theory, then the renormalization of the massless mode is automatically achieved. 

One of the open questions is the underlying connection between the DW fermion and the covariant regularizations. 
While the DW fermion with the PV pairs appears to be a sort of the covariant regularization as in Eq.(\ref{eq: rev_cov}), 
the normalization factor implies that the obtained anomaly is the consistent one, rather than the covariant one. 
Detailed analysis is required in order to understand whether and how the regularization of the DW fermion leads to the chiral gauge theory with no fake anomaly. 
We have discussed it based only on the concrete computation and the observation in the two-dimensional vacuum polarization. 
Through the investigation into the general fermion loops, it is possible to construct a simpler theory without fake anomaly. 
As for the renormalization, in this paper, we have examined the above statements by explicit calculations in the one-loop level.
Although we expect that Eq.\eqref{172017_1Jun18} holds to all orders in the perturbation, it is important to have a general theory of the renormalization in the presence of a space-dependent mass.
Finally, we stress that the PDR is applicable to general field theories. 
It is interesting to apply the PDR to the theories where one has a difficulty in adopting the ordinary dimensional regularization in some sectors.

\subsection*{Acknowledgement}
We thank Y. Kikukawa for valuable comments. 
The work of YH (KS) is supported by the Grant-in-Aid for JSPS Research Fellow, Grant Number 18J22733 (17J02185). 



\end{document}